\documentclass[a4paper,oneside,10pt]{article} 
\usepackage{authblk}

\usepackage{cite}
\usepackage{graphicx}  
\usepackage{dcolumn}   
\usepackage{bm}        
\usepackage{amssymb}   
\usepackage{amsmath}
\usepackage{epsfig}
\usepackage{xspace}
\usepackage{url}
\usepackage{hyperref}
\usepackage{color}
\usepackage{ulem}
\usepackage[utf8]{inputenc}

\newcommand{\bea}{\begin{eqnarray}}
\newcommand{\eea}{\end{eqnarray}}

\newcommand{\Tr}{\textrm{Tr}}

\newcommand{\refeq}[1]{Eq.~(\ref{#1})}

\begin{document}

\title{Topological susceptibility of pure gauge theory using Density of States}

\author[1]{Szabolcs Bors\'anyi}
\author[1,2,3]{D\'enes Sexty}

\affil[1]{ {\it Department of Physics, Wuppertal University, Gaussstr. 20, D-42119 Wuppertal, Germany} }
\affil[2] 
      {\it J\"ulich Supercomputing Centre, Forschungszentrum J\"ulich, D-52425 J\"ulich, Germany}
      
\affil[3] {\it University of Graz, Institute for Physics, A-8010 Graz, Austria}


\maketitle

\date{}

\begin{abstract} 
  The topological susceptibility of the SU(3) pure gauge theory
  is calculated in the deconfined phase
  at temperatures up to $10T_c$. 
  At such large temperatures the susceptibility is suppressed, 
  topologically non-trivial configurations are extremely rare.
  Thus, direct lattice simulations are not feasible. 
  The density of states (DoS) method is designed to simulate rare events,
  we present an application of the DoS method to the problem
  of high temperature topological susceptibility.
  We reconstruct the histogram of the charge sectors that one could
  have obtained in a naive importance sampling. 
  Our findings are perfectly consistent with a free instanton gas.
\end{abstract}

\section{Introduction}\label{sec:introduction}

The past decade has witnessed an immense progress in the theoretical
description of the thermodynamics of strongly interacting matter 
through the advances in the solution strategies of the underlying theory,
Quantum Chromodynamics (QCD). New insights came from a wide range of first
principle approaches ranging from resummed perturbation theory
\cite{Ghiglieri:2020dpq} through functional methods
\cite{Fischer:2018sdj,Gao:2020qsj} to direct simulations on the lattice
\cite{Guenther:2020vqg}.  One of the remaining less understood aspects of QCD
is related to the role of instantons. 
One example is the strong CP problem for which the Peccei-Quinn mechanism
\cite{Peccei:1977hh} offers a solution by introducing the axion particle
\cite{Weinberg:1977ma,Wilczek:1977pj}.  The hypothetical axion is searched for
in various experimental designs, e.g. by shining light through a wall
\cite{Bahre:2013ywa}, helioscopes \cite{Arik:2011rx,Armengaud:2014gea} and
haloscopes \cite{Du:2018uak,TheMADMAXWorkingGroup:2016hpc}. 
The search can be narrowed down by constraints on the axion mass,
e.g by the requirement, that axions have no more contribution to the dark
matter than the total observed abundance
\cite{Preskill:1982cy,Abbott:1982af,Dine:1982ah}. For the latter cosmological
input to be effective we have to obtain information for the axion potential
at the temperatures of the Early Universe, where these were produced.
This strategy was pursued in the framework of lattice QCD in
Ref.~\cite{Borsanyi:2016ksw}.

The QCD axion effectively couples to the gauge invariant but CP breaking
combination of the strong fields
\begin{eqnarray}
\label{chargedef}
q(x)= { 1 \over 32 \pi^2 } \epsilon_{\mu\nu\rho\sigma} 
\Tr (F_{\mu\nu}(x) F_{\rho\sigma}(x))\,.
\end{eqnarray}
This quantity is called topological charge, as its integral $Q$ evaluates to
integer numbers and this charge is topologically stable. Evaluating this term
at temperature $T$ in a space-time volume of $\Omega$ we can express
the topological susceptibility as
\begin{eqnarray}
\chi(T) = \int d^4 x \langle q(x) q(0) \rangle_{T,\Theta=0} = \lim_{\Omega
  \rightarrow \infty} { \langle Q^2  \rangle_{T,\Theta=0} \over \Omega }.
\end{eqnarray}
$\chi(T)$ determines the quadratic term of the axion potential, while higher
order fluctuations of the charge control the details of the shape of the
potential.

The determination of the topological susceptibility using lattice methods
has a long history
\cite{Alles:2000cg,Gattringer:2002mr,Bonati:2013tt,Berkowitz:2015aua,Kitano:2015fla,Borsanyi:2015cka,Bonati:2015vqz,Taniguchi:2016tjc,Petreczky:2016vrs}.
Slightly below the QCD transition temperature susceptibilities close to the
zero temperature value were observed. At high temperatures, on the other hand,
$\chi(T)$ drops with an approximate power law. The power law was actually
expected from the Dilute Instanton Gas Approximation (DIGA) \cite{Gross:1980br}.
The rapid drop of the susceptibility with the temperature manifests in
the finite volume lattice simulations such that calorons (the finite temperature
localized objects that carry a topological charge) are extremely rare.
Lattice QCD simulations would need to sample such rare events with
sufficient statistics to determine at least the variance of $Q$.
The problem of freezing topological sectors in the lattice update algorithms 
poses an additional challenge. Thus, brute force approaches e.g. in
Ref.~\cite{Borsanyi:2015cka} are naturally limited to a short temperature
range.

Several ideas have been proposed to circumvent aforementioned problems. 
In Refs.~\cite{DElia:2013uaf,Bonati:2015sqt} analytic continuation from
imaginary $\Theta$ parameters was used to map out $\chi(T)$ and other
parameters of the free energy, offering a way to calculate higher
moments of the topological charge. Refs.~\cite{Frison:2016vuc,Borsanyi:2016ksw}
addressed the rarity of topological configurations in the Markov chain
of simulation updates. It was observed that at sufficiently high temperatures
configurations with $|Q| \ge 2$ practically never occur even if the
volume is kept large enough to contain the $g^2T$ or even $T_c$ scale. 
Simulating in the $Q=1$ and $Q=0$ sectors separately and determining their
relative free energy provides for an indirect method to calculate $\chi(T)$
at high $T$. While this method was applied successfully, one had to rely
on the cancellation of a quartic divergence in the trace anomaly.
A very different, reweighting based approach was advocated by 
Refs.~\cite{Jahn:2018dke,Jahn:2020oqf} where a modified update algorithm was
introduced to enhance the production of dislocations that may grow into
calorons. Ref.~\cite{Bonati:2018blm} makes a further step and includes
the enhancement force into the microcanonical update rather than deferring
it into a Metropolis step. A common feature of these reweighting based methods
is the use of a proxy charge, which is an easily accessible non-integer
function of the gauge fields that strongly correlates with the integer charge
$Q$. Most of these methods can or have been generalized for the case of
dynamical fermions \cite{Bonati:2015vqz,Borsanyi:2016ksw}.

In many lattice studies configurations with $Q=\pm2$ are mostly missing.
Thus, possible interactions between calorons are not described. While
$Q=2$ configurations have arguably small weight at high temperature,
one may not want to exclude such interactions from the begining.
In Ref.~\cite{Vig:2021oyt} the statistics of both calorons and
anti-calorons were considered (as opposed to the net charge).
The distribution of the topological objects was perfectly consistent
with an ideal gas at a temperature as low as  $T\approx 1.05~\mathrm{MeV}$.

In our work we corroborate the DIGA picture in the high temperature
Yang-Mills theory. We calculate the topological susceptibility using
lattice simulations, not ignoring the very rare $Q=\pm2$ sectors.
We find that the weight of the latter supports the ideal gas description.

We approach the problem of rare calorons with the
Density of States (DoS) method. Originally, the DoS method
studies several small energy ranges separately to determine the
energy dependence of the density of states \cite{Wang:2000fzi}.
The DoS is designed to simulate the physics of rare events.
A prominent example for its successful use is the solution of the sign problem
in heavy dense QCD \cite{Garron:2016noc}.
For a detaled review of the recent progress see \cite{Langfeld:2016kty}
and references therein. This study aims to show for the first time that
the DoS is applicable to the problem of measuring the topological
susceptibility.

In a nutshell, our strategy uses a micro-canonical force on a proxy charge
in the well defined framework of the DoS method. The rare events
with a large proxy charge are sampled and provide for configurations with
$Q=\pm1$ and beyond. As an additional measure to reduce auto-correlation times
we also deploy the parallel tempering method.
We consider here pure gauge theory, expanding the
method to QCD with quark degrees of freedom involves no further
conceptual problems.

In Section \ref{dossec} after a brief overview of the 
DoS method, we explain the details of the application of DoS
to the pure gauge system.
We introduce our lattice setup in Section \ref{setsec} and present
the numerical results in Section \ref{ressec}.
Finally, conclusions are offered in Section \ref{concsec}.

\section{Density of States}
\label{dossec}

First we state the idea of the Density of States method generally, before
we specify to the problem of interest.
Given an action $ S[\Phi] $ of space-time dependent fields $\Phi(x,t)$, we
are interested in the partition function:
\bea
Z = \int D \Phi e^ {-S[\Phi]}.
\eea
Now using the fact that the value of the Gaussian integral
\bea
\int_{-\infty}^\infty dc\ e^{-{P \over 2} (c -a)^2 } = \sqrt{ 2 \pi \over P }
\eea
is independent of the constant $a$, we can rewrite the integral expression
for the partition function such that up to an irrelevant normalization factor we have 
\bea
Z = \int D \Phi  \int_{-\infty}^\infty dc\ e^{ - {P \over 2 } \left( c - F[\Phi]\right)^2 }
e^{-S[\Phi]},
\eea
where $F[\Phi]$ is an arbitrary functional of the fields.
Swapping the order of the integrations we can write
\bea
 Z = \int_{-\infty}^\infty dc\ \rho(c)
\eea
where we defined the 'density of states':
\bea
\rho(c) = \int D \Phi e^{-S[\Phi] - {P \over 2 } ( c - F[\Phi])^2 }.
\eea
If we choose $F$ to be the energy functional and consider the limit
$ P \rightarrow \infty$ than $ \rho(c)$ indeed describes the density of
the energy states of the system, and we get the partition
function as an integral over energy. The formulas remain correct for an
arbitrary functional $F[\Phi]$ and also for finite $P$. In this
case we call $\rho(c)$ the generalized density of states.

We can measure observables using the formula
\bea \label{obsmeas}
\langle A \rangle = {1\over Z } \int_{-\infty}^\infty dc
\int D\Phi A [\Phi] e^{-S[\Phi] - {P \over 2 } ( c - F[\Phi])^2 }
= { \int_{-\infty}^\infty dc\  \rho(c) \langle A \rangle_c \over \int_{-\infty}^\infty dc\ \rho(c)        }
\eea
  where we have defined the notation $ \langle \cdots \rangle_c $, which is an
  average with the action $ S_c[\Phi] = S[\Phi] + {P\over 2}
  (c- F[\Phi])^2 $:
  \bea
  \langle A \rangle_c = {\int D\Phi e^{-S_c[\Phi]} A [\Phi]
    \over \int D\Phi e^{-S_c[\Phi] } }= {1\over \rho(c) } \int D\Phi e^{-S_c[\Phi]} A [\Phi]
 \eea
  
The density of states is reconstructed by measuring the derivative of
its logarithm:
\bea
  { \partial \ln \rho(c) \over \partial c } =
    { 1 \over \rho(c) } \int D \Phi e^{-S[\Phi] - {P \over 2 } ( c - F[\Phi])^2 } \left( -P(c -F[\Phi] ) \right) = \langle -P ( c-F[\Phi]) \rangle_c,
\eea
we can thus measure $\partial_c \ln \rho(c) $ on a predetermined set of points and reconstruct $ \ln \rho(c) $  (and thus $ \rho(c) $) using numerical integration (with e.g. the trapezoid rule). Using this prescription we obtain $ \ln \rho(c) $ with an error magnitude approximately independent of $c$, thus we get $\rho$ with approximately constant relative errors in the whole $c$ range.
An alternative determination
of $\rho(c)$ in the $P\rightarrow \infty $ limit is possible by simply
measuring the histogram of the observable $F[\Phi]$ in a simulation with $P=0$. In this case, however, the statistical errors are proportional
to $ \sqrt{ \rho(c) }$, which can be prohibitive.
Using the former method, thus, allows the determination of the probability
of certain rare events in the configuration space of the theory, which one could not hope to reach in a naive importance sampling simulation. A similar
setup was used in \cite{Fodor:2007vv,Endrodi:2018zda} at nonzero chemical potential $\mu$, where
an additional sign problem is also present. For the present study,
we stay at $\Theta=0$, so the theory has no sign problem.
In a recent paper \cite{Gattringer:2020mbf} the density of states method was
used in a $U(1)$ gauge theory with a $\Theta$-term. There the authors used
open boundary conditions to avoid quantization of the topological charge
and make the system amenable to the DoS treatment. Here we take a different
route, see below.

We will calculate the topological susceptibility 
$\chi=\langle Q^2\rangle/\Omega$,
where $Q$ is the topological charge and $\Omega$ is the four-volume.
At large temperatures in the deconfined phase, the topological
susceptibility is known to be very small \cite{Gross:1980br}.
In an importance sampling simulation
the theory is almost always in the zero charge sector, thus
the value of the susceptibility is given by the probability of rare
visits to the $\pm 1$ charge sectors.
The idea of this paper is to use the density of states method
to measure the probability of these unlikely visits to nonzero charge sectors.
However, the topological charge of the configurations
is given by integer numbers such that the density of states in this case
would be a sum of delta functions on the integer values
(in the $P\rightarrow \infty$ limit), and the procedure described above
is not applicable.
To work around this problem, we need a proxy charge $Q_P[U]$
(a function of the link variables $U$) which
is a continuous value such that $Q_P[U]$ is close to the integer topological
charge $Q[U]$ \cite{Bonati:2018blm}.
Recall the topological charge of a Yang-Mills theory defined
in \refeq{chargedef}.
On the lattice, one can e.g. choose a field theoretical definition of the
charge based on the Wilson flow \cite{Luscher:2010iy,Luscher:2011bx}, similarly
to the
cooling techniques introduced in \cite{deForcrand:1995qq}.
(For other equivalent definitions, see \cite{Alexandrou:2017hqw}.)
One evolves the
gauge field configurations using the flow equations
given by the Wilson plaquette action, and measures a discretised version
of the field strength tensor appearing in \refeq{chargedef}.
One observes that this discretised definition tends to integer numbers
at large flow times, and one can carry out the continuum limit
by fixing the flow time (at which the measurement of the charge is to be
carried out) in physical units.

At zero flow time the gluonic definition of the charge is not close to
integer numbers, and typically it can be
far from the integer topological sector of the configuration. The idea is that if we define the proxy charge $Q_P$ to be the charge
at small flow time, then it will not be restricted to integer numbers.
The integer values are approached after a longer flow time fixed e.g. to the temperature scale.
In order to be able to use a Hybrid Monte Carlo algorithm, we need to be able
to calculate the derivative of $Q_P$ with respect to the gauge fields.
This suggests to use the analytic stout smearing
procedure \cite{Morningstar:2003gk}, so we define
\bea
Q_P[U] =Q_\textrm{clov}[U'_{n,\rho}],
\eea
where $Q_\textrm{clov}$ is the clover discretisation of the topological
charge (\ref{chargedef}) and $U'_{n,\rho}$ are the stout
smeared link variables using $n$ smearing steps with stepsize $\rho$.
A similar use of the proxy charge was described in Ref.~\cite{Bonati:2018blm}.
We thus use $F[U]=Q_P[U]$ in the following to constrain the action.

\section{Simulation setup}
\label{setsec}

The topological susceptibility is often normalized to
the transition temperature's fourth power. On an $N_S^3 \times N_T$ lattice is given by 
\bea
\frac{\chi(T)}{T_c^4}=
    { \langle Q^2 \rangle \over \Omega T_c^4 } = { \langle Q^2 \rangle
      \over ( N_S / N_T )^3 } \left( T \over T_c \right)^4.
\label{voloszt}
\eea

 The gauge action with tree-level Symanzik improvement,
 and the clover discretised
 topological charge density in $Q_P$ is used in the simulations. 

 As discussed in Section~\ref{dossec} a separate simulation must
 be performed for several $c$ values, such that the appropriate range
 of the proxy charge is covered.
 Typically we have used 30-60 $c$ values to measure $\partial_c \ln \rho(c)$ and
 expectation values as a function of $c$. As the theory is symmetric in $c$
 at $\Theta=0$,  we only used non-negative $c$ values, except for a test
 at $T=3T_c$ where we observed good agreement of the results of a
 simulation using $ c \in [ 0,3.5] $ and an other, independent one
 using values $ c\in[-1.2,1.2] $\ ($10^4 \chi /T_c^4 = 1.162(71)\textrm{ vs } 0.98(10)$, respectively).
 We measured the  topological
 charge of every $\sim 50$-th configuration using 
 the gluonic definition after the Wilson flow.
 For the measurement of the exact topological charge $Q$, we use an
 improved discretisation for the topological charge density
 including the $1\times2$ plaquettes in the clover formula
  \cite{Moore:1996wn,deForcrand:1997esx,BilsonThompson:2002jk}, 
which we evaluate at the flow time $t=1/(8T^2) = N_T^2 a^2/8 $.
We round the obtained charge values to integer values, subsequently.

 Unless stated otherwise, results for $N_T=6, N_S=24$ are presented.
 We use Hybrid Monte Carlo for updating the configurations, such that the
 force of the fixing term of the action is calculated in every 3-4th step.
 For the calculation of $Q_P$ we typically use $n=4$ stout smearing steps
 with $\rho=0.1$. The algorithmic parameter $P$ required hand-tuning
 to $P=1000$. Too small values don't constrain
 the dynamics enough to allow extrapolation of $Q\neq 0$ sectors,
 too large values lead to large force terms that require small HMC step sizes
 and thus slow down the simulation.
 
 \begin{figure}
  \begin{center}
  \epsfig{file=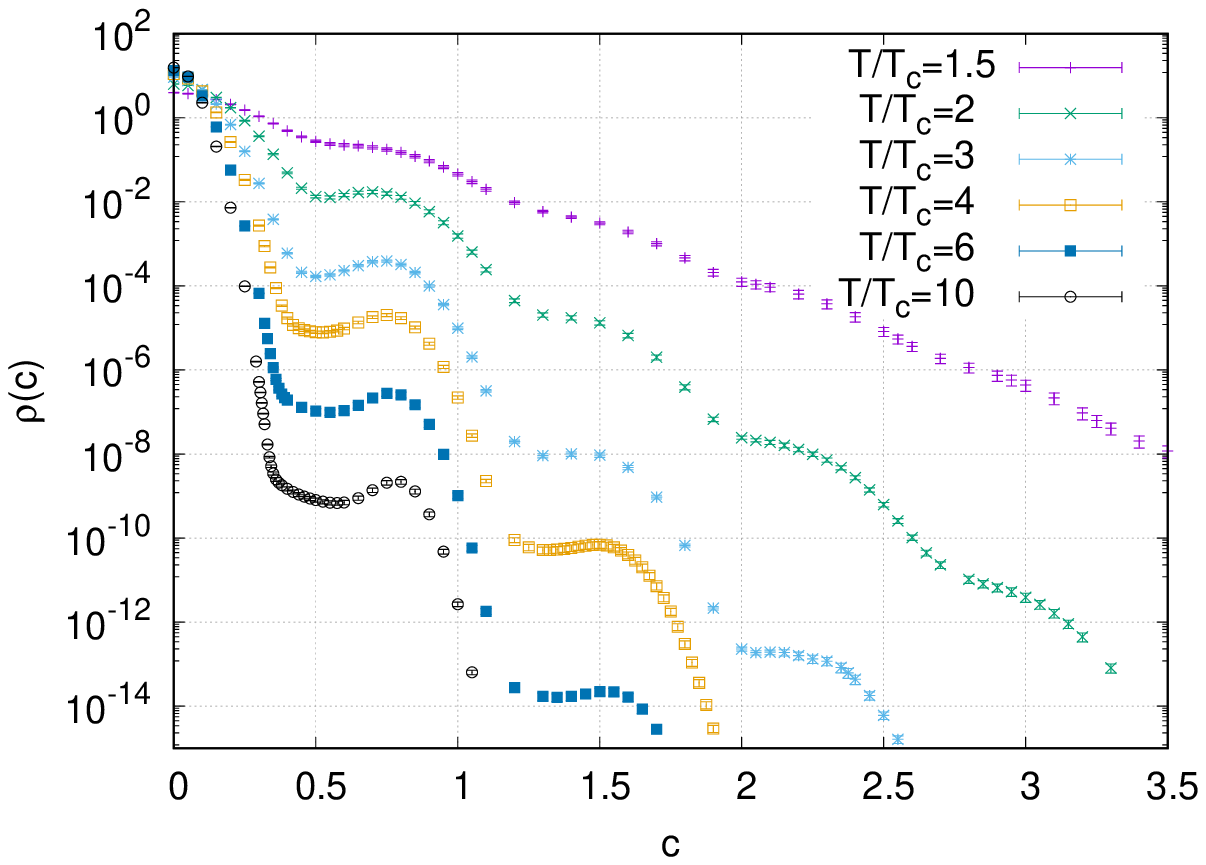, width=8.5cm}
    \epsfig{file=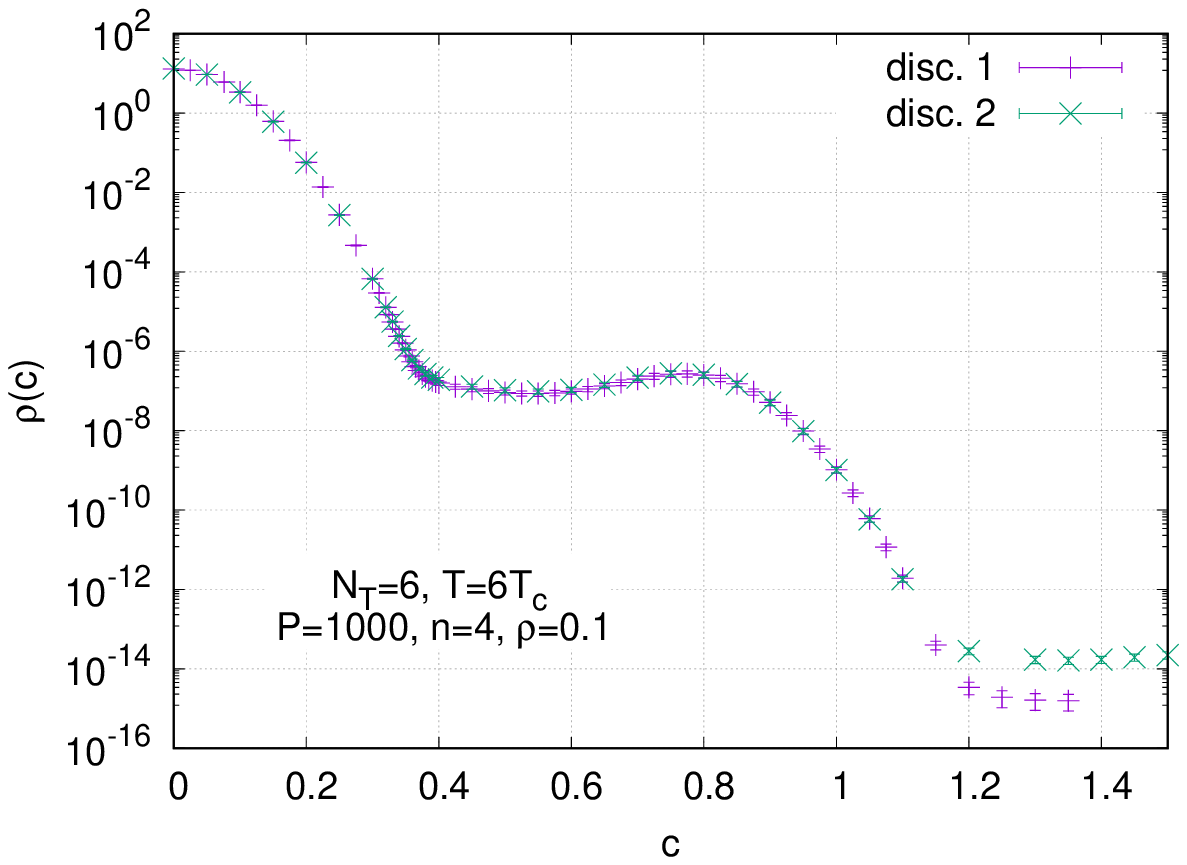, width=8.5cm}
  \caption{     \label{rhofig}
    The density of states for various temperatures (left),
  and at $T=6T_c$ with two different discretisations in $c$ (right).
}
\end{center}
\end{figure}

 Finally $\rho(c)$ is reconstructed by integrating $ \partial_c \ln \rho(c)$.
 Expectation values of observables are calculated from Eq.~(\ref{obsmeas})
 using the trapezoid rule. The undetermined overall factor of
 $\rho(c)$  (which drops out of observables) can be fixed by keeping 
 the integral $ \int_{-\infty}^\infty \rho(c) $ normalized to 1. In practice we
 restrict the integral over positive values of $c$, and the integral
 has a cut-off at the largest $c$ value simulated. 
 Since $\rho(c)=\rho(-c)$ and $\rho(c)$ typically has an overall
 exponential decay for large $c$ this is well justified.
 Statistical errors are
 calculated using the Jackknife procedure. In Fig.~\ref{rhofig} we
 show the reconstructed $\rho(c)$ function for several temperatures. One
 observes a roughly exponential decay for large $c$ which gets faster as the
 temperature is increased.
 At larger temperatures one sees a second local maximum around $ c \sim 0.8
-0.9$, corresponding to configurations which have topological charge $Q=1$. 
At larger $c$ values one can find similar peaks corresponding to the $Q=2,3,\hdots$ sectors.
 
 There is a further ingredient in the used algorithm with the goal
 to reduce auto-correlation times. We use parallel tempering
 \cite{swendsen1986replica,earl2005parallel}
 across the several simultaneously run ensembles, each working on a
 different $c$ parameter.
 Parallel tempering adds a further update step between the 
 HMC trajectories. The update consists of the swap of the gauge
 configurations between ensembles at neighbouring points of the $c$-grid.
 The change in the total action is taken into account by a Metropolis step.
 The tempering update allows dislocations produced at some $c$ to travel
 in the $c$ space and enhance the variety of the configurations at any
 given $c$ parameter. This comes at the price of having correlated 
 errors on the $\rho(c)$ curve. These correlations are correctly kept
 when the $c$-integrals are calculated.

 The grid of $c$ values for a simulation is chosen such that there is
 sufficient overlap between neighbouring simulations. A dense
 grid helps maintaining a high acceptance rate for the tempering updates
 and keeps the systematic error of the integral under control.
Thus, we can keep the systematic error coming from the finite $c$ grid  below the magnitude of the statistical errors. In the 
 right panel of Fig.~\ref{rhofig} two reconstructed
 $\rho(c)$ functions are compared: ``disc.~1'' has twice as many grid points
 as ``disc.~2''. They differ only in the range where $\rho(c)$ drops below 
 $10^{-12}$ as observed in the Figure.
The corresponding $\chi(T)/T_c^4$ values are 
$1.00(18)\cdot 10^{-6}$ and
$1.03(17)\cdot 10^{-6}$, respectively for ``disc.~1'' and ``disc.~2''.
We used the coarser grid for the result plots.

\begin{figure}
\begin{center}
  \epsfig{file=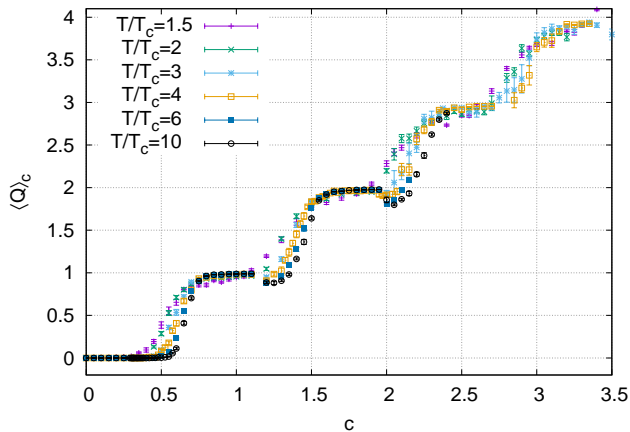, width=8.5cm}
  \caption{ \label{qavrfig}
    The average of the topological charge $Q$ as a function of $c$.
}
\end{center}
\end{figure}

For each $c$ ensemble we determine the $\langle Q\rangle_c$ average,
this we show in Fig.~\ref{qavrfig} for several temperatures.
As expected it roughly follows the $Q=c$ line, and it
has plateaus at integer values: if $c$ is close to an integer
number, then the system stays in the topological sector picked out by $c$.
In the region $ c\approx 0.5 - 0.7$ we see that the average charge
smoothly goes from 0 to 1. By checking the Monte Carlo history of the
topological charge in this region one can see that here the system goes
through many tunnelings and the autocorrelation time of the topological charge
remains small, though with the decrease of the lattice spacing
the situation gradually worsens as expected.

\begin{figure}
\begin{center}
  \epsfig{file=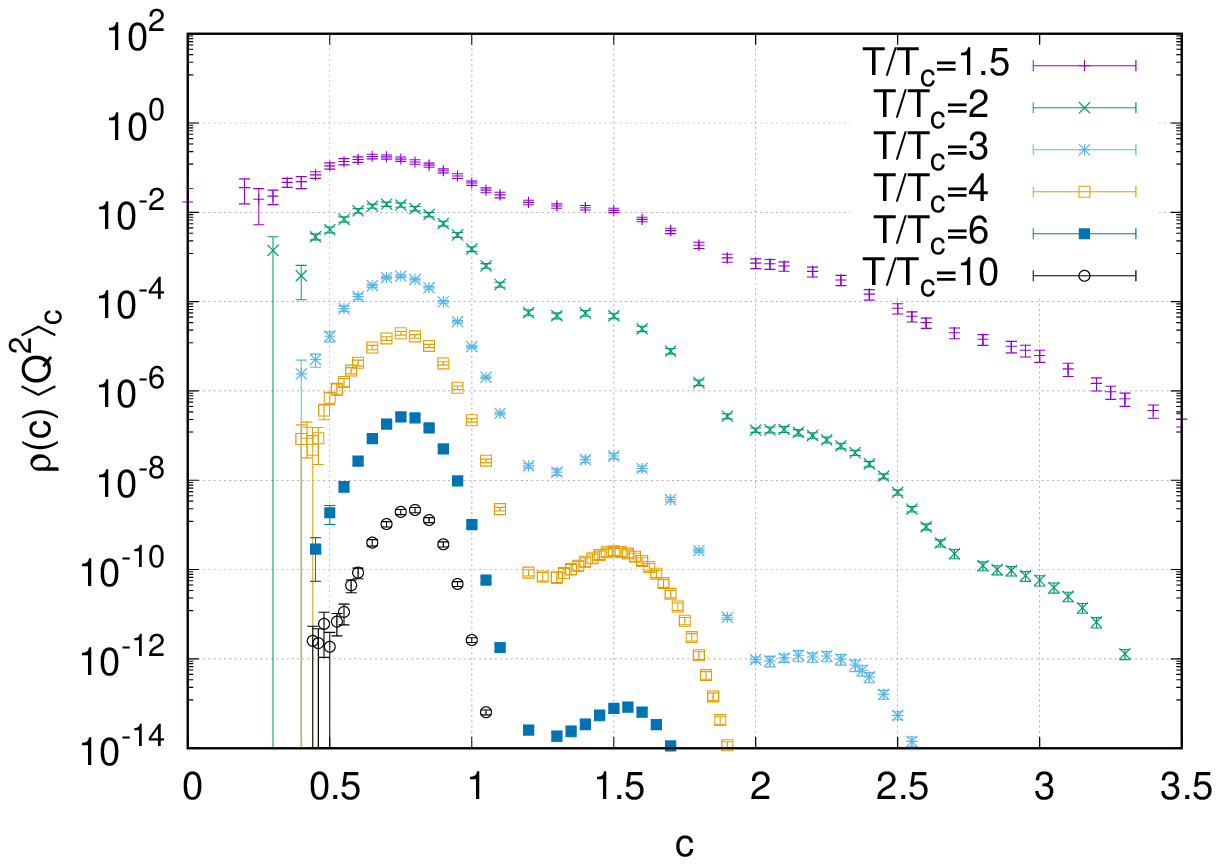, width=6.5cm}
  \epsfig{file=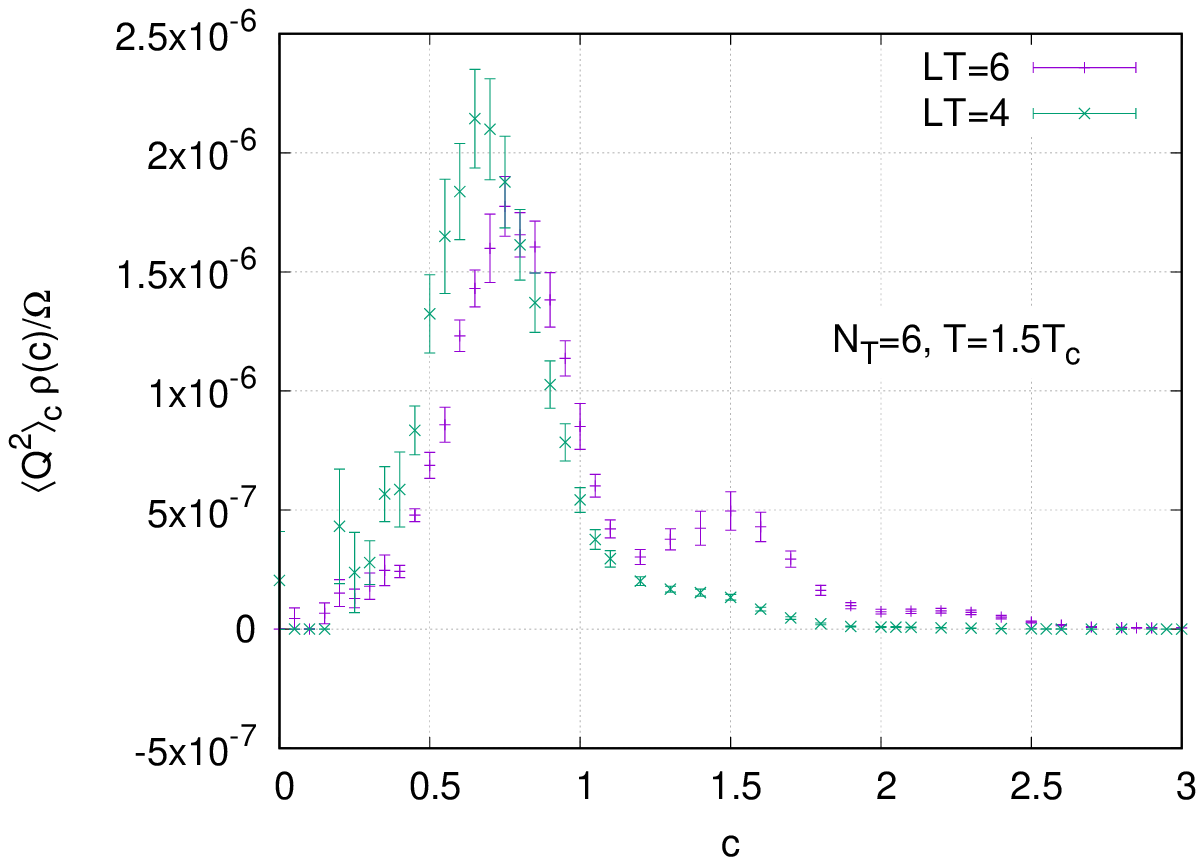, width=6.5cm}
  \caption{ \label{integrandfig}
    The average of the topological charge squared times the density
    of states $\rho(c)$ as a function of $c$.
    The integral of this
    function gives the topological susceptibility.
    On the left panel we show the logarithm of the integrand for various
    temperatures with $LT=4$, on the right panel we fix $T=1.5T_c$ and vary the
    volume in a linear plot.  
}
\end{center}
\end{figure}

In Fig.~\ref{integrandfig} the quantity $ \rho(c) \langle Q^2 \rangle_c  $ is
shown. The integral of this quantity gives the topological susceptibility (we
normalize the integral of $\rho(c)$ to 1). We see that this quantity gets more
and more sharply peaked with increasing temperature, and there the $ |Q|>1$
sectors have negligible contribution to the topological susceptibility. Note
that a logarithmic scale is used which means that even at the smallest
temperature the peak of the $Q=1$ sector carries by far the largest
contribution to $\langle Q^2 \rangle$.
On the right panel we clearly see that the $Q=1$ peak is located at a
value $c<1$, which then translates to a $Q_P$ value significantly less than one.
This was expected since there is a multiplicative renormalization between $Q_P$
and $Q$. After performing the $c$ integral this renormalization factor does not
enter our susceptibility results.

\section{Results}
\label{ressec}

\begin{figure}
\begin{center}
  \epsfig{file=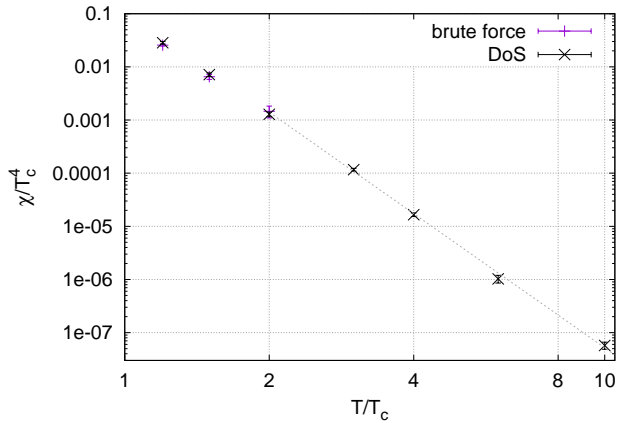, width=8.5cm}
  \caption{ \label{chires}
    The topological susceptibility measured with the brute force
    method and with the DoS approach. We show a power law fit
    with a slope of -6.3(1) with dashed line.
}
\end{center}
\end{figure}

We start with the calculation of the topological susceptibilities
by performing the $c$ integrals at each temperature. 
In Fig.~\ref{chires} we show the resulting $\chi(T)$ function for
$24^3\times 6$ lattices.
Measurements of the susceptibility using direct simulations (i.e. simulations
with $P=0$) are also included for comparison. Direct simulations get
increasingly difficult as the temperature is increased, as in that case the
system is almost always in the $Q=0$ sector, with rare visits to the $ Q=\pm 1$
sector. The quick decay of the probability of configurations with $ Q \neq 0$
makes direct measurements of the topological susceptibility above $ T \sim 4
T_c$ practically impossible \cite{Borsanyi:2015cka}. 

The density of states approach does not depend
on rare tunnelings and can be used at high temperatures, though
at higher temperatures one has to deal with
increasing thermalization and autocorrelation times.
We observe a power law dependence of the susceptibility with the
exponent $-6.3(1)$. The presented $\chi(T)$ as well as the
exponent are affected by discretization effects,
an agreement with the perturbative result \cite{Gross:1980br} is expected to
hold in the continuum limit only.

\begin{figure}
\begin{center}
  \epsfig{file=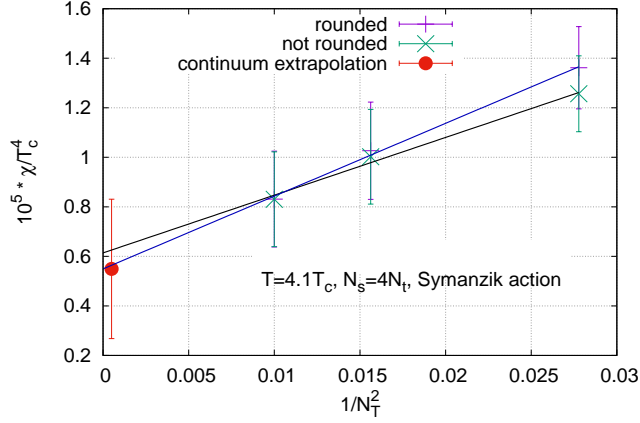, width=8.5cm}
  \caption{ \label{contlimfig}
    Continuum extrapolation of the topological susceptibility
    at $T=4.1 T_c$ using $ N_T=6,\ 8,\ 10$. We show results from the prescription when at flow time $t=1/(8T^2)$ the values of $Q$ are rounded to integer values or when this rounding step is not performed.
  A linear fit for both sets of points is indicated.
 }
\end{center}
\end{figure}

The continuum extrapolation is performed at one point: at $ T=4.1T_c$,
using simulations on $ N_T= 6,\ 8,\ 10,\ N_S=4N_T$ lattices, as visible
in Fig.~\ref{contlimfig}.
The temperature is chosen to facilitate comparison
with the result from \cite{Jahn:2020oqf} which reads: 
$ \chi(T=4.1T_c)/T_c^4 = 4.84 e^{\pm 0.24} 10^{-6} $, where they performed the
continuum extrapolation for the quantity $ \ln \chi $.
(Note that Refs.~\cite{Jahn:2018dke,Jahn:2020oqf} used the plaquette gauge action.) 
Our result is $\chi/T_c^4 =(5.5 \pm 2.8)10^{-6} $ and it's close to the result
we get from continuum extrapolating the logarithm:
$ \chi/T_c^4 = 6.61 e^{\pm 0.30 } 10^{-6} $. The difference of the two
extrapolations one can take as the systematical error of the continuum
extrapolation. One can wonder whether rounding $Q$ to integer values at a
certain flow time has an influence on our results.  In Fig.~\ref{contlimfig} we
show also the values which one gets without the final rounding step. We observe
that
for decreasing lattice spacing the effect of this rounding steadily decreases.
In Tab.~\ref{statisticstable} the number of configurations we used
to calculate the topological charge $Q$ for this continuum extrapolation is
listed. Note that we measured the charge after every $\sim$50 HMC trajectories,
but the quantity $ \partial_c \ln \rho(c)$ is measured more often as it is a
much cheaper observable.

\begin{table}
\begin{center}
  \begin{tabular}{|c|c|c|c|}
    \hline
   $ N_T$  & \# of $c$ vals.& $c$ range & \# of configs. \\
    \hline
    6 & 32 &[0,1.2]& 350 \\
    8 & 32 &[0,1.3]&  688 \\
    10 & 56&[0,1.5]  & 1265 \\
    \hline
  \end{tabular}
\caption{ \label{statisticstable}
  The number and range of the $c$ parameter as well as 
  the statistics used at each $c$ value to calculate the
  continuum extrapolation in Fig.~\ref{contlimfig}.
    }
\end{center}
\end{table}

Next we turn to the question of instanton interactions and the
applicability of the dilute instanton gas approximation (DIGA), which
assumes interactions are negligible.
At large temperatures the appearance of a caloron is
so rare that the probability that two calorons appear close to each other
is small, therefore the DIGA is expected to be a good approximation.  In this case it is expected that the value of $ \langle Q^2 \rangle $ is proportional to  the volume and thus the susceptibility in \refeq{voloszt} is independent of the
volume. 

To investigate this behavior we have performed simulations at $T=1.5T_c$ at two
different spatial box sizes $ L=4/T$ and $L=6/T$. In Fig.~\ref{integrandfig}
(right) the quantity $ \langle Q^2 \rangle_c \rho(c) $ is shown for the two
volumes. Since the $\rho(c)$ function
is normalized to one, the integral of this function gives $\langle Q^2
\rangle$.  One sees that at $LT=4$ mainly the $Q=1$ peak contributes, as in
the smaller volume the appearance of two calorons is relatively rare. In
contrast, in the larger volume the $Q=2$ sector gives a non-negligible
contribution to the susceptibility. The results for the susceptibility are
consistent:
$ \chi/T_c^4 = 0.00716(56) $ on the smaller lattice and 
$ \chi/T_c^4 = 0.00755(48) $ on the larger lattice. 
The $ \chi$ values remain
volume independent, showing that neglecting calorons' interaction is
indeed a good approximation and we could get fairly accurate results
in the smaller volume by restricting our simulations to the $0\le c \le 1.5 $
range.  Because of the strong suppression of the calorons at larger
temperatures the contribution from the higher sectors is even smaller.

We also study the observable $b_2$, defined by
\bea
b_2 = - { \langle Q^4 \rangle - 3 \langle Q^2 \rangle ^2 \over
    12 \langle Q^2 \rangle }
\eea
which characterises the anharmonicity of the axion potential. It is expected
that at large temperatures, where the DIGA approximation holds,
$b_2$ assumes the value -1/12.
In small volumes where only the $Q=0, \pm 1$ sector contributes, 
the value -1/12 follows from the fact that
$ \langle Q^4 \rangle = \langle Q^2\rangle $ and $\langle Q^2 \rangle $
is small. In larger volumes, where calorons and anticalorons appear
independently, their probability distribution follows the
Skellam distribution: $ p_k=e^{-\lambda}I_k(\lambda)$ (see also below), which 
leads to $ b_2= -1/12$ as well.
Earlier results show that starting from $T>1.15 T_c$,
$b_2$ is very close to -1/12 \cite{Bonati:2013tt}. 
In fact in all our simulations $b_2$ is consistent with -1/12 within errors. 
As argued above, this is nontrivial only in the case where $\langle Q^4 \rangle $ has a sizeable contribution from $ |Q| \ge 2 $ sectors.
The simulation using $ LT=6,\ T=1.5T_c $  allows testing this case
and we get a result compatible with the DIGA predicition: $ b_2 = -0.0804 (58) $.

\begin{figure}
\begin{center}
  \epsfig{file=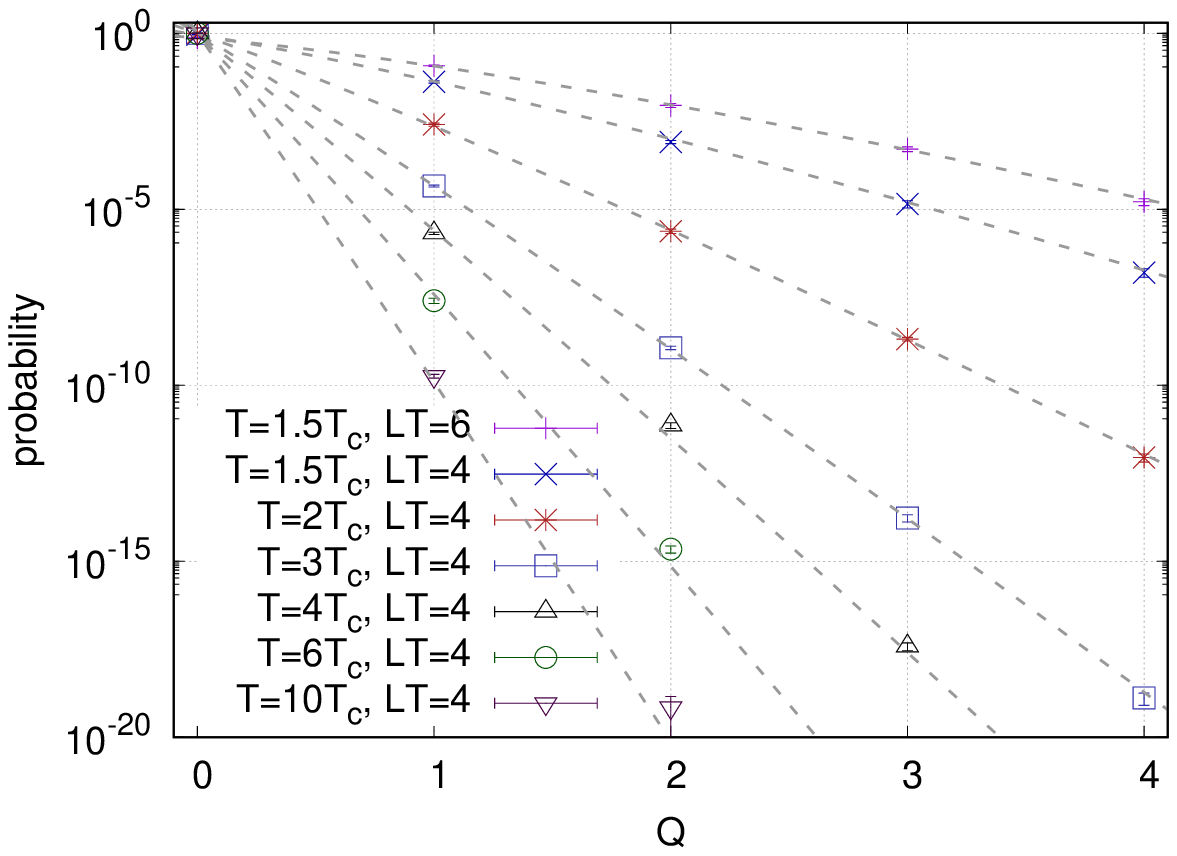, width=8.5cm}
  \epsfig{file=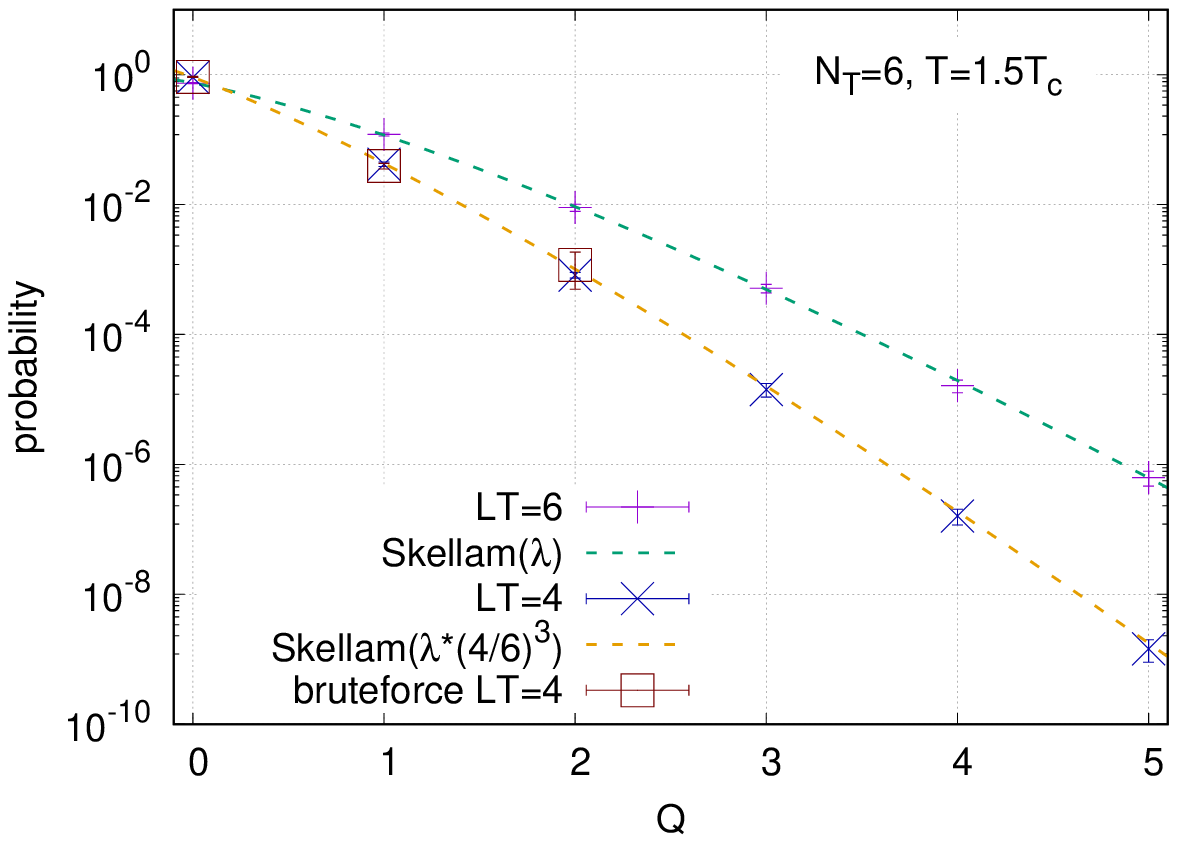, width=8.5cm}
  \caption{ \label{hist}
    The reconstructed histogram of the topological charge in a
    simulation without the forcing term in the action.
    The lines represent fitted Skellam distributions:
    $p_k=e^{-\lambda} I_k(\lambda)$, where the $\lambda$ parameter
    is consistent with $\langle Q^2\rangle$.
}
\end{center}
\end{figure}

The histogram of the topological charge is defined with the help of the
Kronecker-delta function as 
\bea
 h(n)=  \langle \delta_{Q,n} \rangle,
 \eea
 where the overall normalization is chosen such that $\sum_n h(n)=1$.
Using the DoS simulations we can reconstruct this as 
\bea
 h(n)= \int dc \rho(c)  \langle \delta_{Q,n} \rangle_c.
\eea
In Fig.~\ref{hist} we show reconstructed histograms. Note that some of the
probabilities are so low that it would be practically impossible to measure
them using
naive simulations. In the right panel of Fig.~\ref{hist} we show
the reconstructed histograms at $ T=1.5 T_c$ for two different spatial volumes
$ TL=4 $ and $TL=6$. We can model these histograms under the assumption that
instantons are independent of each other: this means
that the number of calorons and anticalorons follow a
Poisson distribution. Thus the total topological charge is distributed
as the difference of two Poisson distributed random variables, known
as the Skellam distribution: $ p_k=e^{-\lambda}I_k(\lambda)$, with
$I_k(\lambda)$, the modified Bessel function of the first kind.
(This distribution is equivalent to takeing
the total number of calorons and anticalorons as a Poisson distributed variable and then
assigning a random sign to each defect \cite{vanderMeulen:2005sp}.)
We observe that the fitted Skellam distribution describes the histograms well.
This verifies that instantons appear independently of each other
in our simulations.
As expected the same $\lambda$ parameter scaled with the volume
describes the different spatial volumes, as one observes
on the right panel of Fig.~\ref{hist}. The results
of a ``bruteforce'' calculation with the unconstrained gauge action are also
shown, using $\approx 3000$ independent configurations.

\section{Conclusions}
\label{concsec}

In this study the Density of States method is applied to the SU(3) pure
gauge theory in order to calculate its topological susceptibility
at high temperatures. In practice, we introduce a force term on a proxy
charge, which may assume non-integer values, but correlates with the
integer topological charge.
This allows the mapping out of the proxy charge density using DoS.

The topological susceptibility is calculated in a temperature range
$T/T_c = 1.2 \dots 10$.
The DoS method also allows reconstructing the histograms of the
topological sectors one would get in a naive importance sampling
simulation. We observe that these histograms follow
the Skellam distribution, which implies that instanton interactions
are negligible at the spatial volumes used in this study.
This is additionally verified by performing simulations at two different
spatial volumes at $ T=1.5T_c$, such that on the larger volume
the $|Q|=2$ charge sector has a non-negligible contribution
to the susceptibility and to the $b_2$ parameter.

In this exploratory study we mostly use $N_T=6$ ensembles, except
for one test at $ T=4.1 T_c$, where the continuum extrapolation is carried out.
The continuum extrapolation over all temperatures is to be carried
out in a follow-up study, allowing a quantitative comparison with
perturbative results. 

Using the method presented here the topological susceptibility at
individual temperatures can be addressed as opposed to the integral
method in Refs.~\cite{Frison:2016vuc,Borsanyi:2016ksw}. Also
the absence of large cancellations between the subtracted free
energy contributions may open the way towards simulations at finer lattices.

For eventual applicability to axion phenomenology,
fermionic degrees of freedom are also required.
Performing a continuum extrapolation with dynamical fermions can be
highly non-trivial and often very fine lattices are required.
Nevertheless, 
since the modified dynamics affects the gauge sector only,
the inclusion of fermions presents no conceptual challenge to the
DoS method described here.


\subsection*{Acknowledgements}

This project was partially funded by the DFG grant SFB/TR55.
We acknowledge funding by
the Gauss Centre for Supercomputing (GCS) for
providing computer time on the JUWELS supercomputer
at the J\"ulich Supercomputing Centre (JSC) under the
GCS/NIC project ID HWU16.
Some parts of the numerical 
calculations were done on the GPU
cluster at the University of Wuppertal, and on the cluster at the University of Graz.

\footnotesize
\bibliographystyle{utphys}

\bibliography{../mybib}
  
\end{document}